\documentclass[apj]{emulateapj}

\newcommand{\myemail}{cmancone@astro.ufl.edu}

\newcommand{\ablue}{-0.97}
\newcommand{\ared}{-0.91}
\newcommand{\errablue}{0.14}
\newcommand{\errared}{0.28}
\newcommand{\mblue}{17.44}
\newcommand{\mred}{17.10}
\newcommand{\errmblue}{0.30}
\newcommand{\errmred}{0.43}

\slugcomment{}

\shorttitle{High-redshift Cluster Galaxy LF}
\shortauthors{Mancone et al.}

\begin{document}


\title{The Faint End of the Cluster Galaxy Luminosity Function at High Redshift}

\author{Conor L. Mancone\altaffilmark{1},
Troy Baker\altaffilmark{1},
Anthony H. Gonzalez\altaffilmark{1},
Matthew L. N. Ashby\altaffilmark{2},
Spencer A. Stanford\altaffilmark{3}$^,$\altaffilmark{4},
Mark Brodwin\altaffilmark{5},
Peter R. M. Eisenhardt\altaffilmark{6},
Greg Snyder\altaffilmark{2},
Daniel Stern\altaffilmark{6},
Edward L. Wright\altaffilmark{7}
}
\altaffiltext{1}{Department of Astronomy, University of Florida, Gainesville, FL 32611}
\altaffiltext{2}{Harvard-Smithsonian Center for Astrophysics, 60 Garden Street, Cambridge, MA 02138}
\altaffiltext{3}{Physics Department, University of California, Davis, CA 95616}
\altaffiltext{4}{Institute of Geophysics and Planetary Physics, Lawrence Livermore National Laboratory, Livermore, CA 94550}
\altaffiltext{5}{Department of Physics and Astronomy, University of Missouri, 5110 Rockhill Road, Kansas City, MO 64110}
\altaffiltext{6}{Jet Propulsion Laboratory, California Institute of Technology, 4800 Oak Grove Drive, Pasadena, CA 91109}
\altaffiltext{7}{UCLA Astronomy, P.O. Box 951547, Los Angeles, CA 90095-1547}

\email{\myemail}

\begin{abstract}
We measure the faint end slope of the galaxy luminosity function (LF) for cluster galaxies at $1 < z < 1.5$ using {\itshape Spitzer} IRAC data.  We investigate whether this slope, $\alpha$, differs from that of the field LF at these redshifts, and with the cluster LF at low redshifts.  The latter is of particular interest as low-luminosity galaxies are expected to undergo significant evolution.  We use seven high-redshift spectroscopically confirmed galaxy clusters drawn from the IRAC Shallow Cluster Survey to measure the cluster galaxy LF down to depths of $M^* + 3$ (3.6$\mu$m) and $M^* + 2.5$ (4.5$\mu$m).  The summed LF at our median cluster redshift ($z=1.35$) is well fit by a \citet{schechter} distribution with $\alpha_{3.6\mu\text{m}} = \ablue \pm \errablue$ and $\alpha_{4.5\mu\text{m}} = \ared \pm \errared$, consistent with a flat faint end slope and is in agreement with measurements of the field LF in similar bands at these redshifts.  A comparison to $\alpha$ in low-redshift clusters finds no statistically significant evidence of evolution.  Combined with past studies which show that $M^*$ is passively evolving out to $z \sim 1.3$, this means that the shape of the cluster LF is largely in place by $z \sim 1.3$.  This suggests that the processes that govern the build up of the mass of low-mass cluster galaxies have no net effect on the faint end slope of the cluster LF at $z \la 1.3$.
\end{abstract}

\keywords{galaxies: clusters: general, galaxies: evolution, galaxies: formation, galaxies: luminosity function}

\section{Introduction}\label{sec:intro}

Many studies have shown that the low-mass cluster-galaxy population evolves substantially at low redshift.  For instance, \citet{cowie96} first recognized that star formation happens primarily in high-mass systems at high redshift and low-mass systems at low redshift, a fact which has been studied extensively since (see, for example, \citealt{panter07}; \citealt{mobasher09}; \citealt{chen09}; \citealt{villar11}).  Moreover, it is well known that cluster galaxies undergo morphological transformation at low redshift, with many cluster members transforming to lenticular galaxies at low redshift \citep{dressler97,desai07,wilman09}.  There is also substantial evidence that the low-luminosity red-sequence galaxy population grows substantially in clusters since at least $z \sim 1$ \citep{stott07,lu09,rudnick09,lemaux12}.

Taken together, these facts demonstrate that the low-mass cluster galaxies are actively evolving and forming since $z \sim 1$.  Therefore, by comparing low mass, $z=0$ galaxies with their high-redshift progenitors we can potentially constrain the processes important in galaxy formation and evolution.  This can be done by studying individual galaxies (through their star formation rates, stellar masses, morphological types, and structural properties) or by studying galaxy populations (through their luminosity and mass functions).

In particular, the near-infrared luminosity function (NIR LF) can be used to study the stellar mass growth of a galaxy population, as the rest-frame NIR is a good proxy for stellar mass \citep{muzzin08}.  In clusters, the NIR LF has been used extensively to study the assembly of the most massive cluster galaxies.  Such studies have found that the massive end of the NIR LF evolves passively out to $z \sim 1.3$, suggesting that the bulk of the stellar mass of these galaxies is in place at high redshift \citep{andreon06,strazzullo06,depropris07,muzzin08,mancone10}.  In addition, in \citet{mancone10} we found statistically significant deviations from passive evolution at $z > 1.3$ which we could only explain with ongoing stellar mass assembly at these redshifts.


\begin{deluxetable*}{ccccc}
\tablecaption{Cluster Member Summary\label{tbl:cluster_summary}}
\tablewidth{0pt}
\tablehead{
  \colhead{Cluster} & \colhead{RA} &  \colhead{Dec} & \colhead{z} & \colhead{\# Members}
}
\startdata
ISCS J1432.4+3332 & 14:32:29.18 & 33:32:36.0 & 1.112 & 26 \\
ISCS J1434.5+3427 & 14:34:30.44 & 34:27:12.3 & 1.238 & 19 \\
ISCS J1429.3+3437 & 14:29:18.51 & 34:37:25.8 & 1.261 & 18 \\
ISCS J1432.6+3436 & 14:32:38.38 & 34:36:49.0 & 1.351 & 12 \\
ISCS J1433.8+3325 & 14:33:51.13 & 33:25:51.1 & 1.369 &  6 \\
ISCS J1434.7+3519 & 14:34:46.33 & 35:19:33.5 & 1.374 & 10 \\
ISCS J1438.1+3414 & 14:38:08.71 & 34:14:19.2 & 1.414 & 16 \\
\enddata
\end{deluxetable*}

Most attempts to probe the faint end of the cluster luminosity function (LF) at high redshift have been limited to studying the red sequence.  Such studies have found a deficit of faint and red cluster members at high redshift when compared to their low-redshift counterparts \citep{delucia04,stott07,lu09,rudnick09,lemaux12}.  This could mean that low-mass cluster galaxies undergo substantial mass growth at low redshift, or simply that low-mass cluster galaxies are still blue at high redshift and have not finished transitioning onto the red sequence \citep{lemaux12}.  Differentiating between these two cases requires measuring the LF of all faint cluster members.  Previously, \citet{strazzullo10} was the only study to do this, finding a faint-end slope consistent with flat.  However, they did not compare their results to low-redshift clusters to determine the implications for the stellar mass growth of low-mass cluster galaxies.

In this paper we measure the 3.6 and 4.5 $\mu$m LF of high redshift ($1 < z < 1.5$) galaxy clusters.  Our measurements trace the rest-frame NIR, where the LF is known to correlate well with stellar mass.  Most importantly, our data are deep enough to constrain $\alpha$, the faint-end slope of the LF.  This, combined with low-redshift results from the literature, allows us to measure the stellar mass buildup of the low-mass cluster galaxy population over a redshift range when these galaxies are known to be actively evolving.

This paper is structured as follows.  Section \ref{sec:data} describes our data.  Section \ref{sec:lf} presents our method for measuring the galaxy cluster luminosity function and gives our results.  In Section \ref{sec:discussion} we compare our results to low-redshift clusters and the field.  Our conclusions are presented in Section \ref{sec:conclusions}.  All magnitudes are on the Vega system, and we assume a WMAP 7 cosmology (\citealt{komatsu11}; $\Omega_m=0.272, \Omega_\Lambda=0.728, h=0.704$) throughout.  All SPS model predictions are generated using {\itshape EzGal} \citep{mancone12}.

\section{Data}\label{sec:data}

\subsection{Cluster Sample}

The clusters from this study are part of the IRAC Shallow Cluster Survey (ISCS) \citep{stanford05,elston06,brodwin06,eisenhardt08}, a catalog of clusters identified as 3-D overdensities using photometric redshifts in the $8.5$~deg$^2$ Bo\"{o}tes field.  Further work with the high-redshift ($1 < z < 1.5$) clusters in the ISCS has included deep (1000s) IRAC imaging, spectroscopic followup, and {\itshape HST} imaging.  Seven of the spectroscopically confirmed, high-redshift ISCS clusters have both deep IRAC imaging and ACS F775W imaging.  It is this subsample of ISCS clusters that we use to study the LF of high-redshift galaxy clusters.  We supplement the 1000 seconds of targeted IRAC observations for each cluster with imaging from the {\itshape Spitzer} Deep, Wide-Field Survey (SDWFS, \citealt{ashby09}) which has a median exposure time of 420 seconds throughout the Bo\"{o}tes field.  This gives a total observing time of roughly 1400s per cluster in all four IRAC bands.

We list in Table \ref{tbl:cluster_summary} our clusters along with their positions, redshifts and number of spectroscopic members.  ISCS J1438.1+3414 has a published X-ray mass estimate of $\log(M_{200}^{L_X}/M_\odot) = 14.35^{+0.11}_{-0.14}$ which comes from a 143 ks Chandra exposure \citep{andreon11,brodwin11}.  All of these clusters (with the exception of ISCS J1433.8+3325) have a weak-lensing mass estimate from \citet{jee11}, with masses in the range of $14.40 \le \log(M^{WL}_{200}/M_\odot) \le 14.73$.

\subsection{Comparison Fields}

\begin{figure*}
\epsscale{1.0}
\plotone{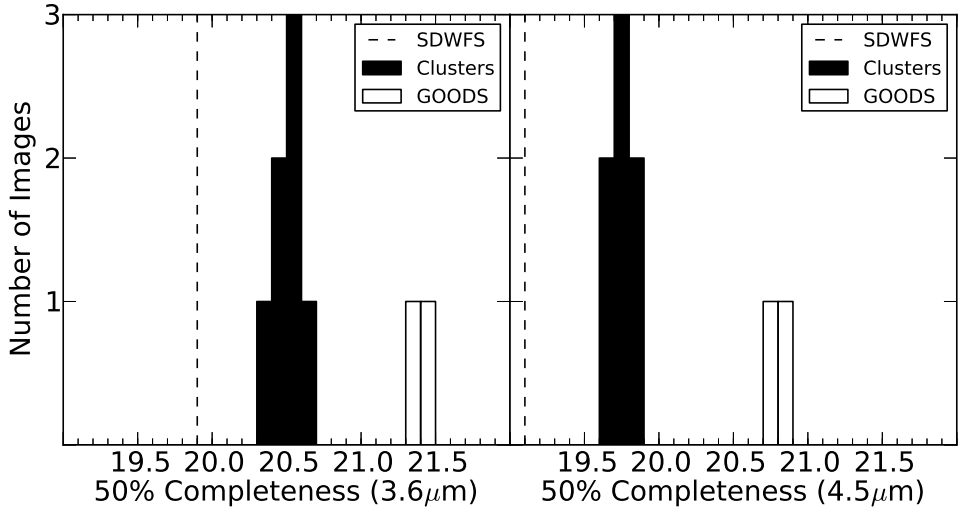}
\caption{Measured 50\% completeness limits for our images in 3.6 (left) and 4.5$\mu$m (right).  In both panels the solid histograms show the completeness limits for our cluster fields, and open histograms show completeness for our control fields.  The vertical dashed line denotes the 50\% completeness limits from the SDWFS survey \citep{ashby09}.}\label{fig:complete}
\end{figure*}

Given the limited spectroscopic redshifts in these fields, statistical background subtraction is required to recover the underlying LF.  A statistical background subtraction involves measuring the number counts of galaxies in a field region and subtracting it from the number counts of galaxies near the cluster.  This technique has been used successfully by \citet{depropris98}, \citet{lin04}, \citet{muzzin08}, and \citet{mancone10}.  This requires a survey with IRAC imaging of at least the same depth as our cluster images as well as ACS F775W imaging.  For this purpose we select the GOODS North and South \citep{goods} fields.  We downloaded the latest fully reduced 3.6 and $4.5\mu$m {\itshape Spitzer} IRAC images taken of the GOODS fields.  We also retrieved the latest {\itshape HST} ACS F775W catalogs from the GOODS survey.  Throughout this paper we refer to the GOODS fields as our control fields.

\subsection{Data Reduction and Processing}\label{sec:reduction}

We produced IRAC mosaics of all seven clusters by combining data from our own programs (PID78, PID30950) with that from SDFWS, following procedures identical to those described in \citet{ashby09}.  This included the manner in which outliers were rejected and in the way the individual IRAC frames were prepared for mosaicing by first removing the residual images arising from earlier exposure to bright sources.  We generated catalogs by running our fully reduced 3.6 and 4.5$\mu$m IRAC images (for the clusters and the control fields) through Source Extractor \citep{bertin96} in single-image mode.  We used 4$\prime\prime$ diameter aperture mags that were aperture-corrected to total mags by comparing 4$\prime\prime$ and 24$\prime\prime$ diameter aperture magnitudes for bright, unsaturated stars in our images.  We used stars to measure the aperture corrections because galaxies at these redshifts are typically unresolved in IRAC imaging.  This gave aperture corrections of $-0.32$ ($-0.34$) magnitudes in 3.6 (4.5)$\mu$m for our cluster images and $-0.31$ ($-0.32$) mags for our control images.  For reference, the difference between the aperture corrections of the cluster and control fields is smaller than the uncertainty of the absolute flux calibration for IRAC images \citep{reach05}.  To verify our calculated aperture corrections we compared our 4$\prime\prime$ aperture corrected magnitudes to the 4$\prime\prime$ aperture corrected magnitudes from SDWFS, and found very small systematic offsets ($<0.03$ mags).

The ACS imaging for our clusters was obtained as part of the {\itshape HST} Cluster Supernova Survey, and the reduction of the images is described in detail in \citet{suzuki12}.  We ran the reduced ACS F775W images through Source Extractor and used MAG\_AUTO to calculate the F775W magnitude of our galaxies.  We used Our F775W photometry to perform an optical$-$NIR color cut to remove contaminants from our LF (Section \ref{sec:color_cut}).  For our control we used the ACS F775W MAG\_AUTO values from the GOODS catalogs, which were also generated with Source Extractor.

Next we calculated completeness as a function of magnitude at 3.6 and $4.5\mu$m for each cluster image and each control image separately.  We approximated our galaxies as point sources due to the coarse IRAC point spread function (PSF).  We generated 24,000 artificial point sources for each image, uniformly distributed between 13 and 25 mags.  Our artificial point sources were simply copies of the PSF for each image, which we generated by median combining unsaturated stars taken directly from each image.  We added these sources to the original images ten at a time, ran Source Extractor again for each new image, and finally calculated the recovery rate as a function of magnitude for a given image and filter.

Figure \ref{fig:complete} shows a histogram of the measured 50\% point source completeness limits for each of our cluster images and control images at 3.6 (left) and 4.5$\mu$m (right).  For comparison the vertical black line denotes the 50\% completeness limit in each band from the SDWFS survey.  We only want to fit for the cluster galaxy LF when the completeness of all galaxies in all clusters is at least 50\%.  Therefore we limit our LF fitting procedure to the the brightest 50\% point source completeness limit for all our clusters, which is 20.37 (19.60) mags in 3.6 (4.5)$\mu$m.  The 50\% completeness limit for our cluster images is $\sim$0.75 mags fainter than for the SDWFS images and $\sim$1 mag brighter than our control images.

\section{Observed Luminosity Function}\label{sec:lf}

\subsection{Optical$-$NIR Color Cut}\label{sec:color_cut}

We use a simple color cut to remove stars and low-redshift galaxies from our catalogs and increase the signal-to-noise ratio of our high-redshift cluster galaxies.  Our color cut is designed to include the blue cloud, as excluding part of it would induce a systematic bias in our measurement of the faint-end slope.  We choose our color cut by using the \citet{bruzual03} stellar models to create a model of a star-forming galaxy with a color on the blue side of the blue cloud, consistent with \citet{lemaux12}.  We then use this same model to estimate the color of the bluest star forming galaxies in our clusters, finding F775W$-$[3.6] $\ge$ 3.5 and F775W$-$[4.5] $\ge$ 3.75.  We use these values for our color cut, and note that our final results are not sensitive to our exact choice because our results change by less than our random errors for a wide range of color cuts ($2.5<\text{F775W}-\text{[3.6]}<5$).

\begin{figure*}
\epsscale{1.0}
\plotone{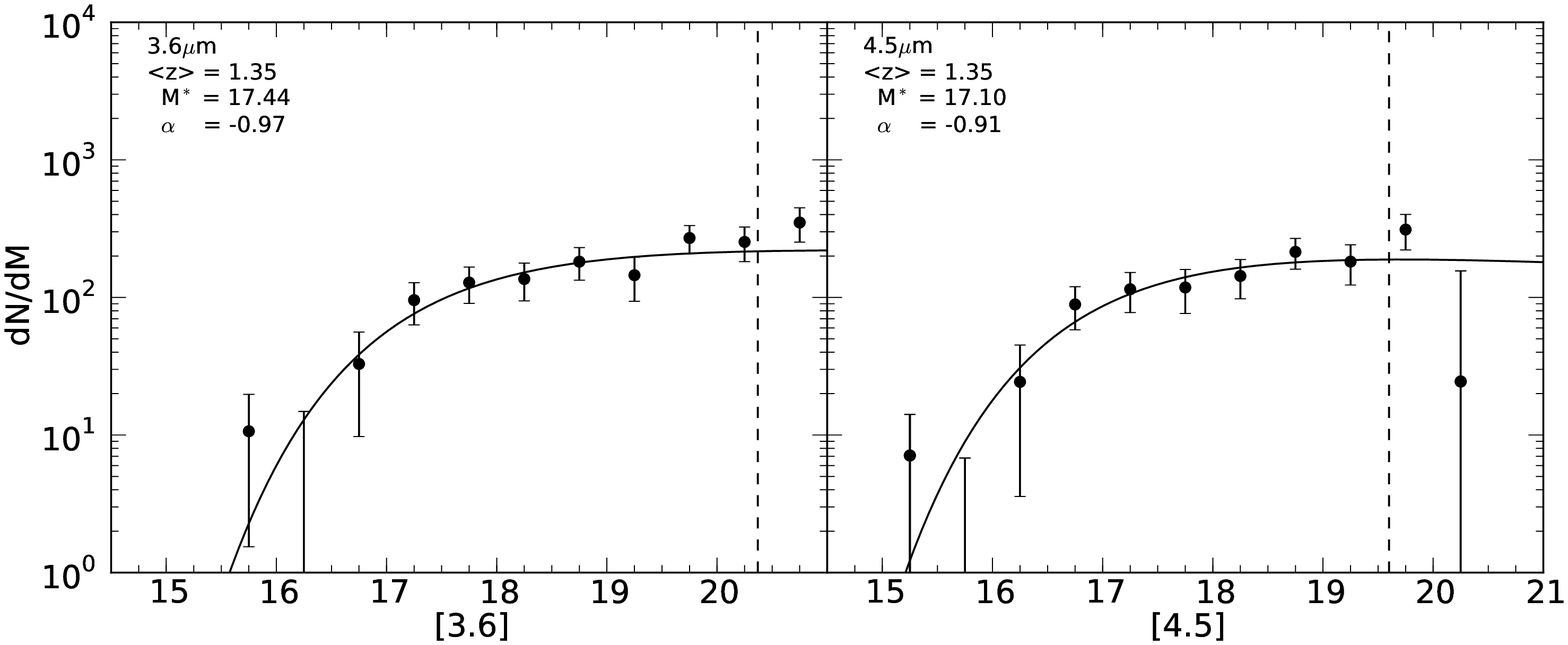}
\caption{Binned and background-subtracted luminosity functions for 3.6 (left) and 4.5$\mu$m (right).  Median cluster redshift and best fitting Schechter parameters are displayed in the top left.  The solid curve shows the best fit.  The dashed vertical line illustrates the magnitude limit for the fit, which is determined by the 50\% completeness limit of the shallowest cluster image.}\label{fig:lfs}
\end{figure*}

Stellar contamination is a potential issue for our sample as our cluster and control fields are at different galactic latitudes.  A [3.6]-[4.5] cut could effectively remove stars from our sample, but is not possible because there is little overlap between the 3.6 and 4.5 $\mu$m imaging of our control regions.  Limiting our sample to areas with 3.6 and 4.5 $\mu$m imaging would remove about 75\% of our control region.  However, stellar contamination is effectively removed via our optical-NIR color cut.  To verify that stellar contamination is not an issue for our results we calculate the expected colors for local stellar populations.  To do this we download stellar isochrones using the CMD 2.3 software\footnote{http://stev.oapd.inaf.it/cmd} which includes the latest stellar modeling details from a number of sources \citep{bonatto04,girardi08,marigo08,girardi10}.  For a low metallicity ($Z=0.008$) model, which is relevant to the Galactic halo, no star of any age has F775W$-$[3.6] $\ga$ 4.  Solar metallicity stars have F775W$-$[3.6] $\sim 5$, at the reddest.  Therefore, only the tip of the RGB and AGB for old stars extend redder than the color cut.  As such, only a small fraction of stars might remain after the cut, meaning that stellar contamination is not an issue.  This is confirmed by the fact that our results do not change even when using color cuts as red as F775W$-$[4.5] = 5, which removes all stellar contamination.  We additionally remove from our catalogs all objects with CLASS\_STAR $>$ 0.8 in the ACS catalogs.  We find that this has a negligible impact on our results, again showing that stellar contamination is not an issue for our sample.

\subsection{LF Fitting Procedure}\label{sec:bgsubtract}

We design our methodology so that we can perform an unbinned fit to the cluster member LF, and so that the background subtraction is done in a way equivalent to a subtraction in observed space.  We generate an individual cluster LF and control LF for every cluster.  The cluster LF is simply composed of the galaxies in the cluster image which passed our various cuts (Section \ref{sec:color_cut}), are within 1.5 Mpc of the cluster center, and are outside of the heavily blended cluster cores (typically $\sim$100kpc).  The latter restriction also removes the BCGs from our sample, which are known to not follow a Schechter distribution.  For each cluster we use a \citet{bruzual03} SPS model to calculate the $k$-correction and distance modulus correction needed to move a passively evolving galaxy ($z_f=3$, Chabrier IMF, solar metallicity) from the cluster redshift to the median redshift of our cluster sample ($z$=1.35).  We then apply this $k$-correction and distance modulus correction to all galaxies in the cluster LF.  Next we build a control LF for each cluster in a similar fashion.  We select all galaxies in the control images which pass our cuts, weight them according to the relative area of the cluster and field images, and apply the exact same $k$-correction and distance modulus correction that we applied to the cluster LF to all the galaxies in the control LF.  We do this so that the same transformation has been applied in the same way to the cluster and control galaxies, and therefore when we subtract the control LF from the cluster LF the subtraction is effectively done in observed space.

This procedure gives us an unbinned cluster and control LF for each cluster.  We then combine the individual cluster and control LFs into a composite cluster and composite control LF, which we use to measure the LF of cluster members.  We parameterize the luminosity function of cluster members as a \citet{schechter} luminosity function and measure the best fitting Schechter parameters with maximum likelihood fitting, similar to the procedure used in \citet{mancone10}.  This procedure requires an analytical representation for the contribution from the control region so we bin our composite control LF by magnitude, correct for photometric incompleteness, and fit a third order polynomial to it in log space.  We then use maximum likelihood fitting to fit the sum of a Schechter luminosity function (the cluster member LF) and the fitted composite control LF to the composite cluster LF.  We use a downhill simplex algorithm \citep{nr} to maximize the likelihood as a function of $\Phi^*$, $m^*$, and $\alpha$, and fit all galaxies brighter than the 50\% completeness limit for the clusters (Section \ref{sec:reduction}).

\begin{figure}
\epsscale{1.0}
\plotone{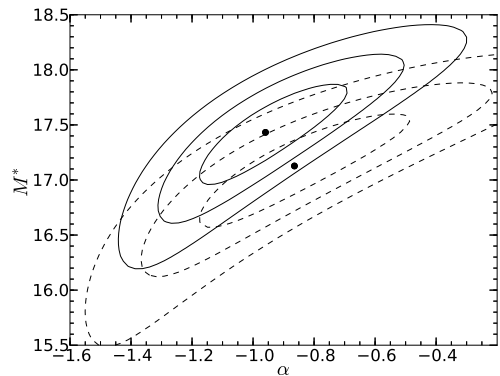}
\caption{Confidence regions for the Schechter fits to our 3.6$\mu$m (filled) and 4.5$\mu$m (dashed) cluster LFs.  Contours represent the 1, 2, and 3$\sigma$ confidence regions in $\alpha$ vs $M^*$ space.  Filled circles denote the best fit Schechter parameters}\label{fig:contours}
\end{figure}

\begin{deluxetable*}{cccccc}
\tablecaption{Best Fitting Schechter Parameters\label{tbl:params}}
\tablewidth{0pt}
\tablehead{
  \colhead{$\langle z \rangle$} & \colhead{\# Clusters} & \colhead{$M^*_{3.6\mu\text{m}}$} & \colhead{$\alpha_{3.6\mu\text{m}}$} & \colhead{$M^*_{4.5\mu\text{m}}$} & \colhead{$\alpha_{4.5\mu\text{m}}$} \\
}
\startdata
1.35  & 7 & $\mblue \pm \errmblue$ & $\ablue \pm \errablue$ & $\mred \pm \errmred$ & $\ared \pm \errared$ \\
\enddata
\end{deluxetable*}

\begin{deluxetable}{ccc}
\tablecaption{Binned LFs\label{tbl:lfs}}
\tablewidth{0pt}
\tablehead{
  \colhead{Mag} & \colhead{\# [3.6]} &  \colhead{\# [4.5]}
}
\startdata
15.25 & $ -1.32 \pm  0.50$ & $  7.09 \pm  7.02$\\
15.75 & $ 10.64 \pm  9.10$ & $ -3.31 \pm 10.12$\\
16.25 & $  0.87 \pm 13.97$ & $ 24.34 \pm 20.76$\\
16.75 & $ 32.86 \pm 23.12$ & $ 88.94 \pm 30.76$\\
17.25 & $ 95.61 \pm 32.39$ & $114.71 \pm 37.00$\\
17.75 & $128.29 \pm 37.82$ & $118.11 \pm 41.72$\\
18.25 & $136.13 \pm 41.68$ & $143.18 \pm 45.37$\\
18.75 & $181.98 \pm 48.57$ & $214.39 \pm 54.09$\\
19.25 & $145.11 \pm 51.32$ & $182.13 \pm 59.06$\\
19.75 & $270.40 \pm 62.50$ &      \nodata      \\
20.25 & $253.25 \pm 71.37$ &      \nodata      \\
\enddata
\end{deluxetable}

\begin{deluxetable*}{cccccc}
\tablecaption{$\alpha$ Values From the Literature\label{tbl:alpha_refs}}
\tablewidth{0pt}
\tablehead{
  \colhead{Reference} & \colhead{Cluster\tablenotemark{a}} &  \colhead{\# Clusters} & \colhead{Band} & \colhead{$\langle z \rangle$} & \colhead{$\alpha$} \\
}
\startdata
\cutinhead{Rest-Frame NIR}
\citet{depropris98}  &            Coma &  1 &         H &    0.023 &         $-0.78 \pm 0.3$\tablenotemark{b} \\
\citet{andreon01}    &          AC 118 &  1 &      K$s$ &      0.3 &        $-1.18 \pm 0.15$\tablenotemark{b} \\
\citet{lin04}        &         \nodata & 93 &      K$s$ &    0.043 &         $-0.84 \pm 0.02$ \\
\citet{jenkins07}    &            Coma &  1 & 3.6$\mu$m &    0.023 &         $-1.25 \pm 0.05$ \\
\citet{muzzin07}     &         \nodata & 15 &         K &    0.296 &         $-0.84 \pm 0.08$ \\
\citet{skelton09}    &           Norma &  1 &      K$s$ &    0.016 &          $-1.26 \pm 0.1$ \\
\citet{depropris09}  &         \nodata & 10 &         K &     0.07 &          $-0.98 \pm 0.2$\tablenotemark{b} \\
\citet{mancone10}    &         \nodata & 35 & 3.6$\mu$m &     0.37 &          $-0.60 \pm 0.2$ \\
This Work            &         \nodata &  7 & 3.6$\mu$m &     1.35 &   $\ablue \pm \errablue$ \\
This Work            &         \nodata &  7 & 4.5$\mu$m &     1.35 &     $\ared \pm \errared$ \\
\cutinhead{Rest-Frame Optical}
\citet{mobasher03}   &            Coma &  1 &         R &    0.023 & $-1.18^{+0.04}_{-0.02}$ \\
\citet{depropris03}  &         \nodata & 60 &        Bj &  $<$0.11 &        $-1.28 \pm 0.03$ \\
\citet{chiboucas06}  &       Centaurus &  1 &         V &   0.0114 &   $-1.4^{+0.1}_{-0.18}$ \\
\citet{strazzullo10} & XMMU J2235-2557 &  1 &         H &     1.39 &   $-1.2^{+0.2}_{-0.15}$ \\
\enddata
\tablenotetext{a}{A cluster name is given only when a single cluster is studied in the given paper.}
\tablenotetext{b}{Formal errors were not given but were estimated from plots of confidence regions.}
\end{deluxetable*}

\subsection{Results}\label{sec:results}

Figure \ref{fig:lfs} shows the control-subtracted cluster LF and the Schechter fit to the cluster member LF for 3.6$\mu$m (left) and 4.5$\mu$m (right).  Maximum likelihood fitting gives a fit to the LF without binning, but for plotting purposes we show the binned and control-subtracted cluster LF in Figure \ref{fig:lfs}, which is the binned difference between the composite cluster LF and the composite control LF.  In each panel the solid curve shows the best fit while the dashed vertical line illustrates the magnitude limit used for the fit.  Figure \ref{fig:contours} shows the 1, 2, and 3$\sigma$ contours in M$^*$ vs. $\alpha$ space derived from our measured likelihoods.  We also report the binned LF values in Table \ref{tbl:lfs}, although we note that our fit was to the unbinned data.

We also measure uncertainties using bootstrap resampling, as such an error estimate is more sensitive to systematic uncertainties caused by cluster-to-cluster variations.  We generate realizations of the LF by randomly selecting seven clusters from our sample and repeating our LF fitting procedure with the new cluster sample.  The cluster selection is done with replacement, which means that an individual cluster can be selected more than once when generating a new cluster sample.  This allows us to probe any systematic uncertainty caused by cluster-to-cluster variation, as this process effectively applies random weights to the clusters while fitting.  We perform 100 realizations of the cluster LF and take the standard deviation of the fitted $M^*$ and $\alpha$ parameters as our random uncertainties.  Our best fitting Schechter parameters and bootstrap errors are listed in Table \ref{tbl:params}.  We note that our measured bootstrap uncertainties agree well with the contours in Figure \ref{fig:contours}.

The relatively small sizes of the GOODS fields means that cosmic variance in our control fields could be an additional source of systematic uncertainty.  To verify that cosmic variance is not strongly biasing our results we redo our fit but use just one of the GOODS fields for our control sample, and then redo it again using the other.  In each case the best fitting value of $\alpha$ changes by $\sim$0.1 in both filters, which is smaller than our measured errors.  Therefore, cosmic variance is unlikely to be a dominant source of uncertainty in our results.

\section{Discussion}\label{sec:discussion}

\subsection{High-Redshift Comparison}

For a basic consistency check we compare to our results from \citet{mancone10}.  In \citet{mancone10} we measured $M^*_{3.6\mu\text{m}}$ and $M^*_{4.5\mu\text{m}}$ out to $z=1.8$ using the slightly shallower SDWFS data and a statistical background subtraction.  As such the methodology is very similar, the filters are the same, and the seven clusters studied herein were also included in \citet{mancone10}.  Due to our shallower data in \citet{mancone10} we fixed $\alpha$ and reported fitted $M^*$ values for $\alpha = -0.6$, $-0.8$, and $-1.0$ in redshift bins from $z = 0.3$ to $z = 2.0$.  The median redshift of the clusters in this study ($z=1.35$) fall directly between the $z=1.24$ and $z=1.46$ bins from \citet{mancone10}.  Therefore we compare our fitted $M^*$ values to the average of the $M^*$ values for the $z=1.24$ and 1.46 bins with $\alpha = -1.0$, which gives $M^*_{3.6\mu\text{m}} = 17.42 \pm 0.1$ and $M^*_{4.5\mu\text{m}} = 16.92 \pm 0.1$, in good agreement with the values of $M^*$ measured herein.

We note that our random errors for $M^*$ in this paper are larger than the quoted errors in \citet{mancone10}.  This is a simple result of number statistics.  While we only have seven clusters in this study, we had 25 (22) clusters in our $z=1.24$ (1.46) bins in \citet{mancone10}.  This leads to a larger random uncertainty in $M^*$ for this current work, although we have lower systematic uncertainty for $M^*$ in this paper because the requirement of fixing $\alpha$ in \citet{mancone10} introduced a systematic uncertainty of $\sim0.2$ mags into $M^*$.

We also compare to \citet{strazzullo10} who measure the $H$-band LF of a $z=1.39$ galaxy cluster to $M^* + 4$.  They find $\alpha_H = -1.2^{+0.2}_{-0.15}$, also consistent with our results.  While there is a difference in passband between our studies, both trace rest-frame wavelengths redward of the 4000\AA~ break so we expect the difference in passband to have a minimal affect on our fitted values of $\alpha$.

Recent studies have found a deficit of low-luminosity red-sequence galaxies in high-redshift clusters \citep{delucia04,rudnick09,lemaux12}.  At face value this seems in contradiction with the flat $\alpha$ values found in this study as well as \citet{strazzullo10}.  However, neither our results nor the results from \citet{strazzullo10} are limited to red-sequence galaxies, and therefore this apparent difference can simply be a sign that low-luminosity galaxies are in place in the cluster environment at these redshifts but have not yet finished transitioning onto the red sequence, as was suggested in \citet{lemaux12}.

\subsection{Low-Redshift Comparison}

We compare our results to low-redshift cluster LFs to assess the evolution of $\alpha$ over a substantial fraction ($\sim$9 Gyr) of cosmic history.  To accomplish this we have compiled a list of $\alpha$ measurements from the literature for a variety of clusters or cluster samples at different redshifts, which we summarize in Table \ref{tbl:alpha_refs}. We note that \citet{depropris98}, \citet{andreon01}, and \citet{depropris09} did not present a formal error for $\alpha$ but did plot confidence regions for their fit, so we estimated the error on $\alpha$ from the plots of their confidence regions.  Specifically, we derived the error from the full range of values covered by their 1$\sigma$ confidence contours.  We split Table \ref{tbl:alpha_refs} up into two sections: studies which trace the rest-frame optical and studies which trace the rest-frame NIR (such as this work).  Star formation can be an important contributor to the rest-frame optical LF, and as such $\alpha$ is not necessarily directly comparable between the two sets of studies.

\begin{figure}
\epsscale{1.0}
\plotone{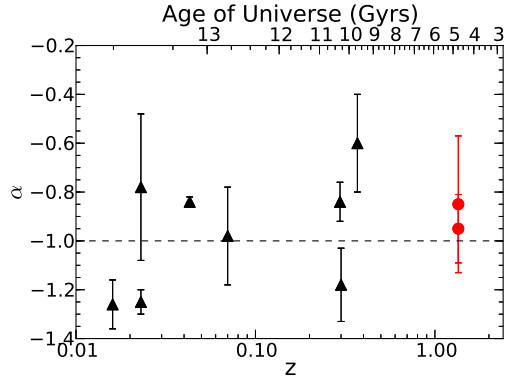}
\caption{Best-fitting values of $\alpha$ to the cluster luminosity function versus redshift.  Data points come from the rest-frame NIR studies from Table \ref{tbl:alpha_refs} and include a variety of literature results.  The results from this work are shown as red circles while previous results are shown as black triangles.  The dashed horizontal line denotes $\alpha=-1.0$, corresponding to a flat LF for faint galaxies.}\label{fig:z_alpha}
\end{figure}

The results in Table \ref{tbl:alpha_refs} are presented graphically in Figure \ref{fig:z_alpha}.  In this Figure the fitted $\alpha$ values and errors are plotted for all the rest-frame NIR results in Table \ref{tbl:alpha_refs}.  There is substantial study-to-study scatter at low redshift, and large error bars at high redshift, but Figure \ref{fig:z_alpha} shows no obvious evidence for evolution in $\alpha$ from $z = 0$ to $z \sim 1.4$, representing nearly 70\% of cosmic history.

Past work on cluster LFs have primarily characterized the evolution of $M^*$, and shallow imaging has required assuming a value for $\alpha$ and fixing it as a function of redshift (see, e.g., \citealt{muzzin08} and \citealt{mancone10}).  Fixing $\alpha$ has been a potential source of systematic uncertainty, as the strong coupling between $M^*$ and $\alpha$ means that if $\alpha$ is improperly held fixed then the fitted values of $M^*$ will also be wrong.  This can be particularly important for studies of the evolution of $M^*$ because if $\alpha$ is evolving but assumed to be fixed then this false assumption can create spurious evolution in $M^*$.  This potential source of systematic uncertainty was discussed in detail in \citet{mancone10} because we found that for $z\gtrsim1.3$ the fitted values of $M^*$ to the cluster LF deviated strongly from passive evolution.  We concluded that while evolution in $\alpha$ could contribute to the measured deviations from passive evolution, it was unlikely to be the underlying cause because the direction of the deviation would require $\alpha$ to become steeper at higher redshift.  Having now measured $\alpha$ at high redshift we can conclude that evolution in $\alpha$ was not the cause of our observed deviations from passive evolution as $\alpha$ does not evolve significantly with redshift out to $z \sim 1.4$.

\subsection{Comparison to the Field LF}\label{sec:field_comparison}

We find that at high redshift the faint-end slope of the cluster LF matches that of the field.  \citet{saracoo06} measured $\alpha$ in the rest-frame $J$-band for field galaxies, finding $\alpha = -0.94^{+0.16}_{-0.15}$ in a redshift bin centered at $z \sim 1.2$.  \citet{cirasuolo07} found $\alpha = -0.92 \pm 0.18$ for galaxies with $1.25 < z < 1.5$ in the rest-frame $K$-band, again in good agreement with the faint-end slope of the cluster LF found herein.  Moreover, work in the field \citep{saracoo06,cirasuolo07,stefanon12} has also found that the faint end of the LF for field galaxies is consistent with being constant out to the highest redshifts studied.  For example, \citet{kochanek01} find $\alpha = -1.09 \pm 0.06$ for 2MASS galaxies at low redshift, while \citet{stefanon12} find that the faint-end slope of their rest-frame $J$-band LF is consistent with flat from $1.5 < z < 3.5$.  They compare their results to lower redshift studies, finding no evidence for evolution in $\alpha$ out to $z = 3.5$ with a mean value of $\alpha = -1.05 \pm 0.03$.  This is all consistent with our finding that $\alpha$ does not substantially evolve but is consistent with flat from $z = 0$ to $z \sim 1.4$ for cluster galaxies.

\subsection{Implications for Galaxy Formation}

As discussed above, we find no statistically significant evidence for evolution in the faint-end slope of the NIR luminosity function out to $z \sim 1.4$.  A lack of evolution in $\alpha$ combined with a lack of evolution in $M^*$ \citep{mancone10} means that both faint and bright galaxies are largely in place at high redshift.  This places a strong constraint on the luminosity evolution of cluster galaxies at $z \lesssim 1.4$.  Either little evolution is happening at lower redshifts, or the processes responsible for LF evolution have no net impact on the cluster population.

To understand the implications of this for galaxy evolution, we must understand how the various process that cause cluster galaxy evolution would affect the luminosity evolution of the cluster galaxy population.  The fact that the NIR traces old stellar populations means that our results are most sensitive to processes which would cause evolution in the stellar mass of cluster galaxies.  Any process which affects low and high stellar mass cluster galaxies equally will lead to evolution in $M^*$.  Conversely, any process which leads to differential evolution between galaxies with low and high stellar masses will lead to evolution in $\alpha$.

The two primary processes by which galaxies can grow their stellar masses over time are star formation and mergers.  The downsizing paradigm (see \citealt{fontanot09} and references therein) suggests that star formation will be preferentially found in lower mass galaxies at low redshift.  Such a mass dependence for galaxy star formation histories will necessarily imply
evolution in the faint-end slope of the LF.  The amplitude of this effect however depends upon the total amount of ongoing star formation, which will vary between clusters and may depend upon the total mass of the host cluster halo.  There is evidence that substantial star formation is still ongoing in our cluster sample \citep{snyder12}, as well as other clusters at similar redshifts \citep{hilton10,tran10,fassbender11}.  In contrast, \citet{muzzin12} find that star formation has already been strongly quenched in their cluster at $z=1.2$.

Mergers can also build up the stellar mass of cluster galaxies, although mergers are expected to be suppressed in the cluster environment due to the high relative velocities of cluster galaxies \citep{alonso12}.  Recent theoretical studies \citep{murante07,conroy07,puchwein10} suggest that for massive galaxies growth by mergers becomes very inefficient (but see \citealt{rudnick12}), and it is possible that mergers can yield a steepening of the faint-end slope if this efficiency is strongly mass-dependent.

In contrast gravitational interactions such as galaxy-galaxy interactions, interactions of a galaxy with the cluster potential, galaxy harassment, and the dissolution of cluster galaxies, can all strip mass away from low-mass galaxies in particular \citep{moore96,boselli06,murante07} and therefore cause $\alpha$ to grow flatter or turn over with time.

Another process for consideration is the infall of new galaxies into the cluster, the effect of which depends on the shape of the LF for the infalling galaxy population.  Since we find that the cluster and field galaxy populations have a similar faint end slope (Section \ref{sec:field_comparison}), we expect that the infall of new galaxies into the cluster will primarily act to mitigate any potential evolution of the cluster LF by driving the cluster LF back towards a flat faint-end slope.  Instead, the infall of new galaxies will lead to an increase in $\Phi^*$, the normalization of the LF, which we have not constrained

Finally, a mass-dependent galactic initial mass function (IMF) can cause the shape of the LF to change relative to the underlying stellar mass function.  This is because the rate of luminosity evolution for a stellar population depends sensitively on the IMF \citep{conroy09}.  Recent work \citep{cappellari12} has suggested that the IMF does indeed depend on galaxy mass, such that lower mass galaxies have a flatter IMF (i.e., a higher fraction of high-mass stars).  A flatter IMF leads to a faster fading of the underlying stellar population \citep{conroy09} and therefore, if true, the results of \citet{cappellari12} suggest that low-mass galaxies should fade faster than high mass galaxies when the stellar masses of both remains fixed.  This will cause $\alpha$ to grow flatter or turn over with time, effectively acting against processes which build up the stellar mass of galaxies but without impacting the underlying mass function.

Clearly, there are many processes which could potentially cause evolution of the NIR cluster LF at $z < 1.5$.  Therefore, the lack of evolution in $\alpha$ observed herein, combined with the lack of evolution in $M^*$ observed over a similar redshift range \citep{mancone10}, places an important constraint on these processes.  In net, they cannot cause any large evolution in the shape of the NIR cluster LF.  This could be because both low mass and high mass cluster galaxies are largely assembled at high redshift, or because the differing effects of these processes causes little evolution in net.  The latter might imply an uncomfortable degree of fine tuning.  In general though, the ability of infalling galaxies to dilute any evolution in the cluster LF allows for more flexibility in the strength of other processes.

\section{Conclusions}\label{sec:conclusions}

We measure the 3.6 and 4.5$\mu$m luminosity functions of seven galaxy clusters at $1 < z < 1.5$, specifically investigating the shape of the LF for faint galaxies.  We find the LFs to be well-fit by Schechter distributions with faint-end slopes of $\alpha_{3.6\mu\text{m}} = \ablue \pm \errablue$ and $\alpha_{4.5\mu\text{m}} = \ared \pm \errared$, both consistent with having flat faint-end slopes within 1$\sigma$.  Our primary conclusions are summarized here:

\begin{enumerate}
\item
We compare to studies of the NIR LF of low-redshift clusters and find no statistically significant evidence for evolution of the faint-end slope of the cluster LF.  Therefore we conclude that the faint end of the cluster LF has not evolved significantly over 70\% of cosmic history.
\item
Having measured a non-evolving faint-end slope we have removed one source of systematic uncertainty from studies of the evolution of $M^*$ as a function of redshift.  This is particularly relevant for our recent detection of deviations from passive evolution at high redshift \citep{mancone10}.  Shallow imaging in \citet{mancone10} necessitated fixing $\alpha$, which could have lead to spurious evolution in $M^*$ if $\alpha$ was evolving.
\item
We compare to the faint end slope for field galaxies at similar redshifts and find good agreement.  Field studies \citep{saracoo06,cirasuolo07,stefanon12} find a faint-end slope consistent with flat at high redshift, and the most recent results \citep{stefanon12} find no evidence for evolution out to $z=3.5$.
\item
Given recent studies \citep{muzzin08,mancone10} which have found that the evolution of the bright end of the cluster LF is consistent with passive evolution out to $z \sim 1.3$, we conclude that the shape of the cluster LF has been in place and evolved little since $z \sim 1.3$.  This could suggest that low-mass galaxies are largely assembled at high redshift.  Conversely, it could simply mean that the many processes which cause evolution of the cluster galaxy population have no net impact on the mass and luminosity function of cluster galaxies.
\end{enumerate}

\acknowledgments
This work is based on observations made with the Spitzer Space Telescope, which is operated by the Jet Propulsion Laboratory, California Institute of Technology under a contract with NASA.  The observations are associated with programs P78 and P30950.  This work is based on observations made with the NASA/ESA Hubble Space Telescope, obtained from the Data Archive at the Space Telescope Science Institute, which is operated by the Association of Universities for Research in Astronomy, Inc., under NASA contract NAS 5-26555.  The observations are associated with program GO-10496.  CLM and AHG acknowledge support for this work from the National Science Foundation under grant AST-00708490.  AHG, MB, and SAS also acknowledge support from NASA through grant HST-GO-10496.

\bibliographystyle{apj}
\bibliography{galaxy_lf}

\end{document}